\documentclass[twoside]{article}
\usepackage{fleqn,espcrc2,amsfonts,amssymb}
\usepackage{graphicx}
\usepackage{cbdef}

\renewcommand{\setZ}{{\mathbb Z}}

\renewcommand{\setR}{{\mathbb R}}

\title{Wavelet-induced renormalization group for the Landau-Ginzburg model}
\author{C. Best\address{John von Neumann Institute for Computing,
        52425 J{\"u}lich, Germany}}

\newcommand{\A}{{\cal A}}
\renewcommand{\AA}[1]{{\cal A}^{(#1)}}

\begin{document}

\begin{abstract}
The scale hierarchy of wavelets provides a natural frame for
renormalization. Expanding the order parameter of the
Landau-Ginzburg/$\Phi^4$ model in a basis of compact orthonormal
wavelets explicitly exhibits the coupling between scales that leads to
non-trivial behavior. The locality properties of Daubechies' wavelets
enable us to derive the qualitative renormalization flow of the
Landau-Ginzburg model from Gaussian fluctuations in wavelet space.
\end{abstract}
\maketitle

\section{INTRODUCTION}

Daubechies' wavelets\cite{Daub,Chui} are an orthonormal basis that explicitly
separates scales. The basis functions are generated from a
single set of functions $\psi_t(x)$ 
by dyadic dilatations and translations. 
Each basis function is expressed as
\be
  \psi^{(n)}_t(x')(x) = 2^{-nD/2} \psi_t(2^{-n}(x-x')) \quad,
\ee
where $n\in\setZ$ is the scale, $x'\in{\cal L}^n$
the position on a grid ${\cal L}^n = (2^n\setZ)^D$ with spacing $2^n$,
and $t=1,\ldots,n_t=2^D-1$ determines how the $D$-dimensional
wavelet is composed of one-dimensional functions. The
resulting wavelet fields naturally live on lattices ${\cal L}^n$ 
of a dyadic multigrid.

It was proved only by Daubechies that a compact basis with these
features (and good analytical properties) exists.  It appears a
natural basis to perform renormalization in.  In this contribution, we
demonstrate how a wavelet basis can be used to exhibit explicitly the
interscale coupling introduced by the $\Phi^4$ term and how an
approximation to the renormalization flow in the
Landau-Ginzburg/$\Phi^4$ model can be derived from this.

\section{WAVELET EXPANSION OF THE $\Phi^4$ THEORY}

\begin{figure}
\includegraphics[width=\hsize]{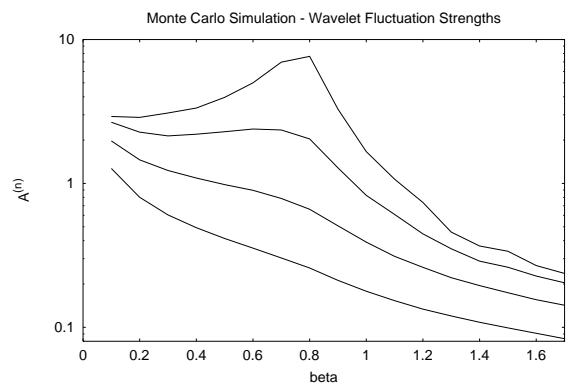}
\vspace*{-1.5cm}
\caption{Wavelet fluctuation strengths at different scales
as a function of the inverse temperature $\beta$,
measured in a Monte Carlo simulation\label{figMc}}
\end{figure}

We consider the Landau-Ginzburg/$\Phi^4$ theory \cite{Goldenfeld}
with a one-component
real order parameter $S(x)$, $x\in\setR^D$, governed by the
Hamiltonian
\be \label{eq2}
  {\cal H} = \int{\rm d}^Dx \left[
      \frac{1}{2} \left(\nabla S(x)\right)^2
     +\frac{r_0}{2} S(x)^2
     +\frac{u_0}{2} S(x)^4 \right]
\ee
with coupling constants $r_0$ and $u_0$. The field is expanded in a
wavelet basis:
\be \label{eq1}
  S(x) = \sum_n \sum_t^{n_t} \sum_{x'\in{\cal L}^n}
         S^{(n)}_t(x') \psi^{(n)}_t(x')(x)
         \, + S_0
\ee
Since wavelets have vanishing first moments, we use the real number $S_0$
to represent the overall magnetization of the system, while the
$S^{(n)}_t(x')$ represent fluctuations at different scales.
Fig.~\ref{figMc} shows their strength $\left\langle
\left[S^{(n)}_t(x')\right]^2 \right\rangle$ as measured in a Monte
Carlo simulation. As they show very clearly the location of the phase
transition, we will focus on modelling these quantities.

We thus make the variational ansatz that fluctuations are Gaussian and
diagonal in wavelet space:
\be \label{eq3}
  \langle S^{(n_1)}_{t_1}(x'_1) S^{(n_2)}_{t_2}(x'_2) \rangle
  = \delta_{n_1,n_2} \delta_{t_1,t_2} \delta_{x'_1,x'_2} \,
    {\cal A}^{(n_1)}_{t_1}
\ee
Note that local fluctuations in wavelet space still provide 
for a nontrivial correlator in position space, as the decay of the
correlator is encoded in the relative strength of fluctuations at
different scales $n$.

The magnitude of the fluctuations ${\cal A}^{(n)}_t$ is then found
by the variational method
from minimizing the free energy ${\cal F} = U - {\cal S}/\beta$
where the internal energy $U$ is the expectation value of the
Hamiltonian $H$ in the Gaussian ensemble (\ref{eq3}), and ${\cal S}$
the entropy of this ensemble.

When calculating the internal energy, the self-similarity of the
wavelet basis comes into play: As all basis functions are built from a
single mother wavelet, the matrix elements of the Laplace operator
have a simple scaling form:
\bea
  \lefteqn{
    \int{\rm d}^Dx \, \psi^{(n_1)}_{t_1}(x'_1)(x) \Delta
                      \psi^{(n_2)}_{t_2}(x'_1)(x)} 
  \nnm \\ &&
   = 2^{-2n} C_{t_1,t_2} \quad.
\eea
Similarly, the four-point overlap
integral occurring from the $\Phi^4$ term has an
approximate scaling representation. In this approximation,
the effective internal
energy per site reads simply
\bea
  \frac{U}{N_0} &=&
  \frac{1}{2} \sum_{nt} 2^{-n(D+2)} (-C_{tt}) \, \AA{n}_t
   \nnm\\ && 
  +\frac{r_0}{2} \A
  + \frac{3u_0}{2} \, \A^2 + 3 u_0 \bar S^2 \A \nnm\\&&
  + \frac{r_0}{2} \bar S^2 + \frac{u_0}{2} \bar S^4
\eea
($\bar S = \langle S_0 \rangle$) with the fluctuation sum
\be
  {\cal A} = \sum_n \sum_t 2^{-nD} \AA{n}_{t} \quad.
\ee
while the Gaussian terms are linear in the fluctuation strengths, the
$\Phi^4$ term introduces a coupling between scales. Since wavelets
have compact support, the overlap integrals and thus the nonlinear
contribution is finite, as opposed to the situtation with a Fourier
basis.  The canonical dimension $2$ of the Laplace operator enters in
the weight factor of the first term.

\begin{figure}
\includegraphics[width=\hsize]{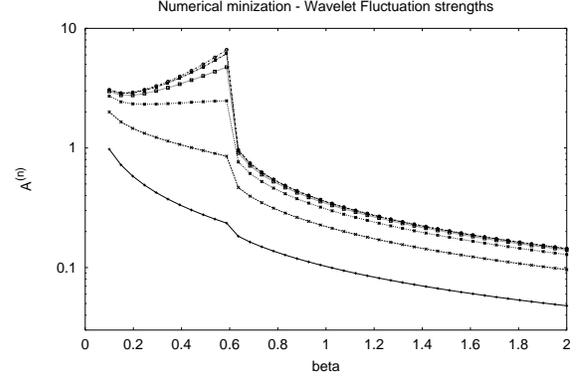}
\vspace*{-1.5cm}
\caption{Wavelet fluctuation strengths at different scales
as a function of the inverse temperature $\beta$,
calculated from minimizing the effective internal energy.\label{fig1}}
\end{figure}

Minimizing the free energy now with respect to the magnetization $\bar S
=\langle S_0 \rangle$
yields
\be \label{eq4}
  \bar S = 0 \qquad\mbox{or}\qquad
  \bar S = \sqrt{-\frac{r_0}{2u_0} - 3 \A } \quad.
\ee
Thus spontaneous symmetry breaking occurs when the fluctuation sum $\A$
exceeds the critical value $-r_0/6u_0$.
Similarly, minimizing with respect to the $\AA{n}$ results in
\be \label{eq5}
  \AA{n}_t 
  = \frac{1}{\beta} \, \frac{1}{
      \frac{1}{2} \, 2^{-2n} (-C_{tt})
      + \frac{1}{2} r_0 + 3u_0 (\A + \bar S^2) 
    } 
\ee
This is an implicit equation for the fluctuation sum $\A$ that can 
be solved numerically. Fig.~\ref{fig1} shows such a solution.

For $u_0=0$, (\ref{eq5}) would be the wavelet representation of a
exponentially decaying correlator with inverse correlation length
$\sqrt{r_0}$.  We can see that the interaction leads to a redefinition
of the inverse correlation length depending on the quantity $\A$ which
is defined implicitly by this equation.  In particular, the
correlation length diverges at the same point as the second solution
in (\ref{eq4}) becomes real, signifying the phase transition.

\section{RENORMALIZATION FLOW}

\begin{figure*}[tb]
\vspace*{-0.5cm}
\begin{tabular}{ll}
\begin{minipage}[t]{.77\hsize}
\hbox to\hsize{\hfil}
\includegraphics[width=\hsize]{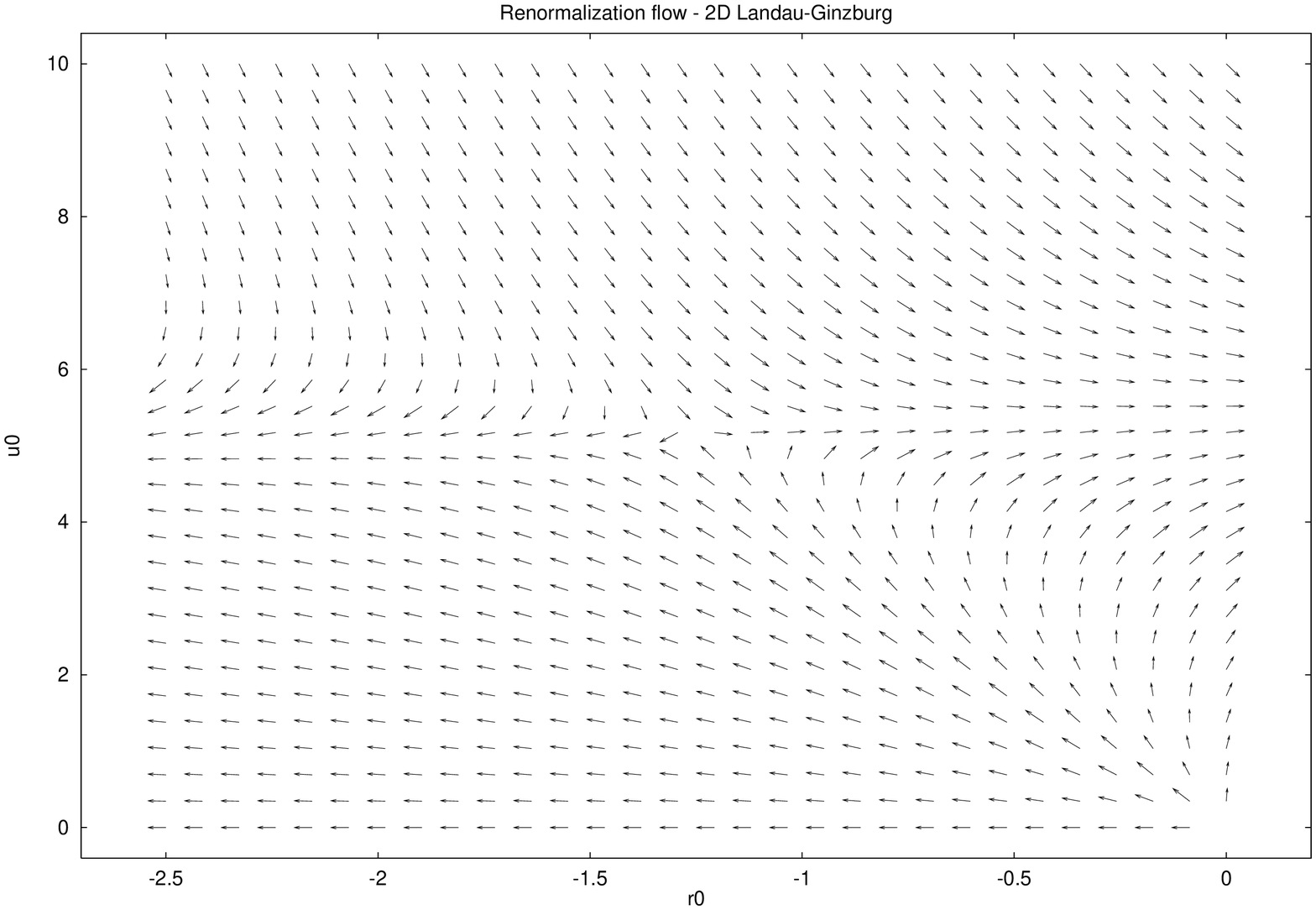} 
\end{minipage}
&
\begin{minipage}[t]{.22\hsize}
\hbox to\hsize{\hfil}
\vspace{-0.8cm}
\caption{%
Renormalization flow of the effective Wavelet Hamiltonian in two dimensions.
Shown is the $r_0>0$, $u_0<0$ quadrant of the coupling-constant plane.
The arrows indicate the direction of the renormalization map.
The Gaussian fixed point is at the lower right corner, the second
fixed point at the center. To the right would be the symmetric phase,
to the left the symmetry-broken phase.}
\end{minipage}
\end{tabular}
\end{figure*}

To derive the renormalization group flow of the theory, we apply the
idea of removing the fine degrees of freedom to the
minimization of the free energy, i.e., we perform the minimization
scale by scale.

Assume there exists a lowest scale $n=0$, e.g.~by already having
eliminated all scales finer than $n$. We can then minimize the free
energy with respect to $\AA{0}_t$, keeping all coarser scales as
variables. $\AA{0}_t$ will then become a function of the $\AA{n}_t$,
$n>0$. By reinserting this function in the expression for the free
energy and expanding in a Taylor series, one reaches a new effective
free energy, now only depending on $\AA{n}_t$, $n>0$,
in which the new terms arising from the expansion can be
absorbed (to some order) in a redefinition of the coupling constants.
This yields a renormalization flow as a mapping 
$(r_0,u_0) \to (r'_0,u'_0)$ in the coupling constant plane. 
It turns out that there
exists the Gaussian fixed point $r_0=u_0=0$, and in $D<4$ a second
fixed point, corresponding to the Wilson-Fisher fixed point.

The actual position of the fixed points still depends on the matrix
element of the Laplacian $-C_{tt}$ and thus on the type of wavelet.
This shows a limitation of our approximation as we disregarded the
correlation between neighboring wavelets which is basically determined
by the extent of the wavelet.

\section{CONCLUSIONS}
 
A wavelet expansion can be used to derive the properties of the
Landau-Ginzburg model and its nontrivial renormalization flow even in
a rather simple approximation.  The crucial features we have made use
of are scaling and self-similarity of the basis and locality of the
basis functions.  They enabled us to focus on the fluctuation
strengths at different scales as the quantities of interest that
govern the phase transition.  The effective free energy of the system
exhibits in a minimal way the coupling between different scales.

\end{document}